\documentclass[preprint,12pt]{elsarticle}

\usepackage{epsfig}
\usepackage{graphicx}

\usepackage{amssymb}

\journal{NIM A}

\begin{document}

\begin{frontmatter}



\title{Evaluation of SiC detector performances for energy and timing measurements}

\author[label1]{N.S. Martorana}
\affiliation[label1]{organization={INFN-Sezione di Catania},
             city={Catania},
            country={Italy}            
            }
            
\author[label1,label2]{G. D'Agata}
\affiliation[label2]{organization={Dipartimento di Fisica e Astronomia "Ettore Majorana"},
             city={Catania},
            country={Italy}            
            }

\author[label1,label2]{A. Barbon}

\author[label1]{G. Cardella}
\author[label1]{E. De Filippo}
\author[label1,label2,label3]{E. Geraci}
\affiliation[label3]{organization={CSFNSM},
            city={Catania},
            country={Italy}
            }
            
\author[label4]{C. Guazzoni}
\affiliation[label4]{organization={DEIB Politecnico Milano and INFN Sez. Milano},
           city={Milano},
           country={Italy}
            }
            
\author[label5,label6,label7]{L. Acosta}
\affiliation[label5]{organization={Instituto de F\'isica, Universidad Nacional Aut\'onoma de México},
           city={Mexico City},
           country={Mexico}
            }   
\affiliation[label6]{organization={Instituto de Estructura de la Materia, CSIC},
           country={Madrid, Spain}
            }            
                   
\author[label7]{C. Altana}
\affiliation[label7]{organization={INFN-LNS},
           city={Catania},
           country={Italy}
            }  
            
\author[label4]{A. Castoldi}

\author[label1,label2,label3]{A. Composto}

\author[label7]{S. De Luca}

\author[label7]{P. Figuera}

\author[label1,label2]{B. Gnoffo}

\author[label8]{F. La Via}
\affiliation[label8]{organization={Institute for Microelectronics and Microsystems (IMM), National Research Council (CNR)}, 
            city={Catania}, 
            country={Italy}
}

\author[label7]{C. Maiolino}

\author[label7]{E.V. Pagano}

\author[label1]{S. Pirrone}

\author[label1,label2]{G. Politi}

\author[label1,label9]{L. Quattrocchi}
\affiliation[label9]{organization={Dipartimento MIFT, Università di Messina},
           city={Messina},
           country={Italy}
            } 
            
\author[label1,label9]{F. Risitano}    
        
\author[label2,label3,label7]{F. Rizzo}

\author[label7]{P. Russotto}

\author[label7]{G. Sapienza}

\author[label1,label9]{M. Trimarchi}

\author[label7]{S. Tudisco}

\author[label2,label3,label7]{C. Zagami}

\begin{abstract}
The development of new detectors based on Silicon Carbide (SiC) is currently a topic of interest within the scientific community. The significant features of SiC make it highly promising for detecting charged particles, neutrons, and $\gamma$/X radiation. In this framework, within the SAMOTHRACE (Sicilian Micro and Nano Technology Research and Innovation Center) ecosystem, an array of new-generation SiC detectors is under development, specifically designed for nuclear and medical investigations using radioactive ion beams. This paper describes the results obtained in the characterization of SiC prototypes regarding  energy and timing measurements. A new method, based on coincidence data analysis, is employed to evaluate the timing performances of SiC detectors. The obtained results have been compared with tests performed using a micro-channel plate as a start detector reference for timing measurements.

\end{abstract}




\end{frontmatter}


\section{Introduction}
Silicon Carbide (SiC) is a well-known wide band-gap semiconductor, currently employed in various technological fields, ranging from aerospace applications, renewable energy, medical and environmental devices, as well as nuclear research. The main focus of the present work is the potential development of SiC-based charged particle detectors \cite{M.DeNapoli_10.3389/fphy.2022.898833,S.Tudisco_s18072289}.\\ 
SiC has a tetrahedral structure,  where each silicon (Si) atom is bonded to four carbon (C) atoms. SiC exists in more than 200 polytypes, classified by their lattice symmetry as cubic (C), hexagonal (H), or rhombohedral (R). Each polytype is further identified by a numerical prefix that denotes the number of layers in its stacking sequence. The most common and commercially available polytypes -- 3C-SiC, 6H-SiC, and 4H-SiC -- present high thermal stability, enabling the growth of large ingots. Among these, 4H-SiC is the preferred choice for detection applications due to its wider band gap. Furthermore, 4H-SiC has several features that make it advantageous compared to other detectors, such as the Si and diamond-based ones \cite{M.DeNapoli_10.3389/fphy.2022.898833,S.Tudisco_s18072289}. For example, the high band gap provides low noise and visible-light blindness, the high displacement energy and thermal stability ensure radiation hardness and efficient heat dissipation, while the high electron saturation velocity provides a low charge trapping probability and a fast signal \cite{M.DeNapoli_10.3389/fphy.2022.898833,S.Tudisco_s18072289}. The latter is critical for time-of-flight (TOF) technique and other nuclear applications where high time resolution is needed \cite{M.DeNapoli_10.3389/fphy.2022.898833}. In addition, SiC-based detectors show good performance in terms of energy resolutions and a linear response with the energy deposition \cite{ref:TUDISCO2025170112}. Furthermore, advancements in the production of SiC detectors have now reached a good accuracy level. Therefore, the use of SiC is nowadays a good compromise between the known reliability of silicon detectors and the radiation hardness properties of diamond ones \cite{C.Altana_s23146522}.\\ 
The interest in the use of SiC--based detectors is driven by both nuclear and medical physics that involve high intensity stable and radioactive ion beams (RIBs). Experiments with high-intensity stable and RIBs are considered essential for the future of nuclear physics \cite{ref:Martorana_frontiers}. 
The former would be important to study low reaction cross-sections, while the latter would allow to study nuclei far from the stability valley, addressing important topics such as probing of nuclear structure, clustering, isospin physics and collective modes, and to explore the study of reactions cross-sections of astrophysical interest involving unstable nuclei (as the r- process) \cite{ref:Martorana_frontiers, ref:Agodi_midtermplanlns, ref:midterm_LNL}.\\ 
Additional interest stems from medical physics and the potential development of SiC-based dosimeters, micro-dosimeters, and beam monitors \cite{G.Parisi_10.3389/fphy.2022.1035956}. These devices enhance clinical radiation therapy by enabling precise dose measurements, crucial for effective treatment and patient safety. Additionally, their high radiation hardness makes them suitable for new techniques-like the FLASH radiotherapy \cite{G.Parisi_10.3389/fphy.2022.1035956}. In this medical context, RIBs also offer innovative possibilities for targeted cancer therapy \cite{ref:MDurante, ref:DBoscolo, ref:boscolo2024}. 
Hadronterapy, as proton therapy, benefits from the Bragg peak effect, ensuring a favorable depth-dose distribution in tissue \cite{ref:boscolo2024}, and even more, therapy with accelerated $^{12}$C-ions, adds biological advantages. In contrast, particle therapy is more sensitive to uncertainties in the beam range than conventional X-ray therapy. To address this, techniques such as positron emission tomography (PET) detect $\beta^{+}$-emitted by isotopes like $^{11}$C and $^{10}$C, formed via $^{12}$C beam fragmentation in the patient’s body. However, PET with $^{12}$C ions remains limited in reducing range uncertainties to the desired levels, due to the relatively low production rate of $\beta$ emitters. A promising alternative is the use of RIBs, such as $^{11}$C, which could enhance the treatment efficacy \cite{ref:MDurante}. Despite their potential, RIB-based therapies require advanced production facilities, specialized detectors for event-by-event ion characterization, and further clinical studies to optimize therapeutic protocols.\\
Within this framework, new worldwide facilities are in construction or in upgrading to produce high-intensity RIBs \cite{ref:Martorana_frontiers, ref:Agodi_midtermplanlns, ref:midterm_LNL, ref:SPES, ref:ISOLDE_facility, ref:GANIL_facility, ref:FRIB_MSU, ref:FAIR_GSI, ref:Russotto_2018}, giving an important boost to perform  medical investigations as well as for the future of nuclear physics.\\
In this context, many efforts have been made aiming at developing new-generation SiC detection systems 
\cite{S.Tudisco_s18072289,C.Altana_s23146522, ref:Martorana_frontiers,ref:Martorana_IWM2024}. The final goal of our project is to develop an array of SiC devices, able to operate in harsh environments with RIBs and capable to measure energy and time, with a resolution of 1$\%$ and better than $\sigma_{t}$ $\approx$ 100 ps, respectively. A high timing resolution for SiC detectors (117 $\pm$ 11 ps FWHM) has been already measured in ref. \cite{ref:Zhang_6520962}, using small size pixel detectors of 400 $\times$ 400 $\mu$m$^{2}$. Our ultimate goal is to reach similar sub-nanoseconds timing resolutions by using much larger size SiC detectors. For the specific timing performance, it is important to underline that developments of dedicated electronics front-end is crucial to allow the optimization of the timing performance of these detectors. For this specific task, a new fast electronics front-end, in development at the Politecnico and Sezione INFN di Milano, will be coupled to the final detection device \cite{ref:Martorana_frontiers,A.Castoldi_10141686}.\\
This paper discusses first results obtained in the characterization of new-generation SiC detectors using radioactive $\alpha$-sources and low-energy proton and $\alpha$ beams, with a focus on their performances for energy and timing measurements. In particular, a new method used to evaluate the timing performance of detectors is presented. These results have been compared with the outcomes of a test performed using a micro-channel plate (MCP) detector as a start signal for the timing measurement. 
\section{Experimental setup and characterization using radioactive $\alpha$ sources}
\begin{figure}
\centering
\includegraphics[scale=0.90]{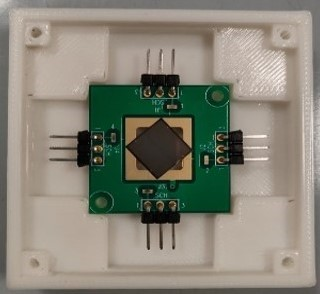}
\includegraphics[scale=0.90]{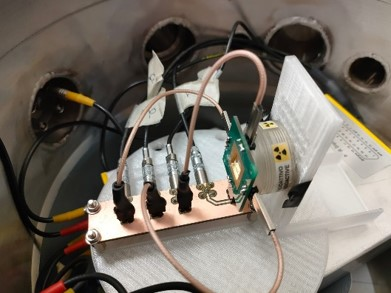}
\caption{(left) Picture of 2$\times$2 pixels SiC detector. (right) Picture of mixed $\alpha$-source placed in front of the SiC detector.\label{Picture_SiC_detector}}
\end{figure}

\begin{figure}
\centering
\includegraphics[scale=0.30]{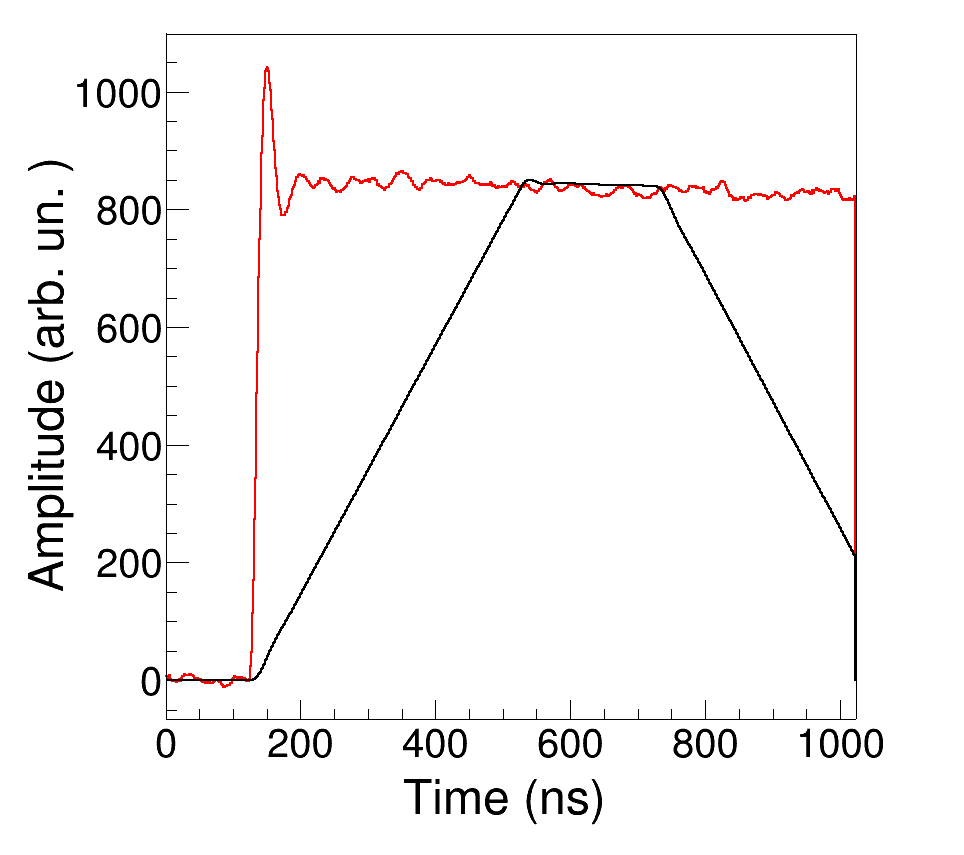}     
\caption{Red line: SiC waveform obtained using a mixed $\alpha$ source in vacuum. Black line: signal after the applied trapezoidal filter used to extract the maximum.\label{figure_waveforms}}
\end{figure}

\begin{figure}
\centering
\includegraphics[scale=0.50]{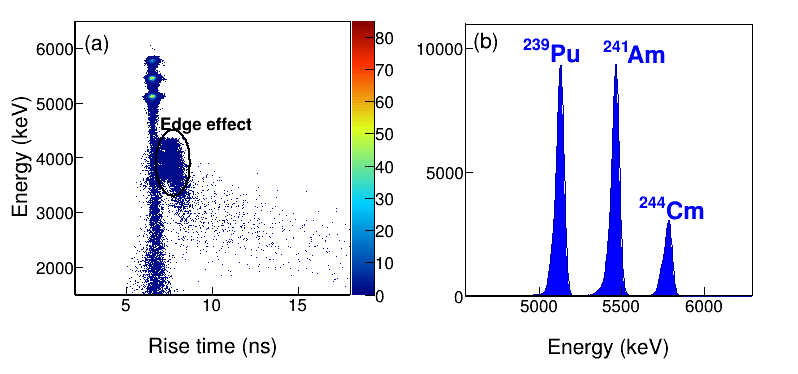}
\caption{(a) Energy vs rise-time plot obtained using a mixed $\alpha$-source, for a SiC pixel. (b) Energy calibrated spectrum obtained with a cut in M$=$1 and in rise-time $\lesssim$ 7 ns. \label{energy-risetime-spectra}}
\end{figure}

\begin{figure}
\centering
\includegraphics[scale=0.5]{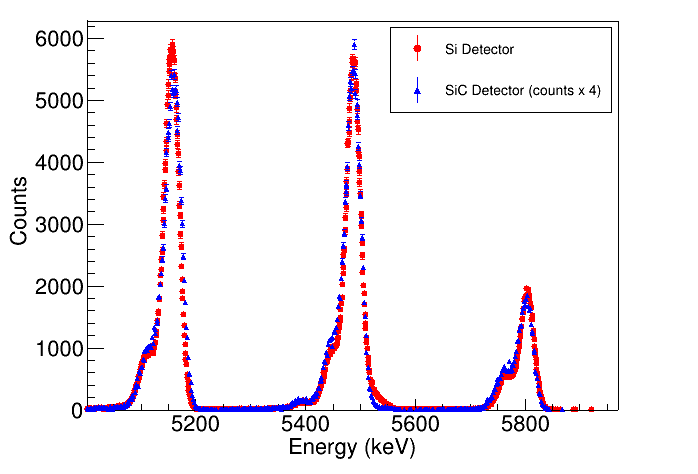}
\caption{Blue dots: Energy spectrum obtained with a Mesytec preamplifier, an ORTEC amplifier and a MCA analyzer for a 100 $\mu$m SiC pixel. Red dots: same energy spectrum for a 100 $\mu$m Si detector \label{MCA-spectrum}.}
\end{figure}

The characterization has been performed using a 2$\times$2 pixels SiC detector (100 $\mu$m thick, 1 cm$^{2}$ of surface and a depletion voltage of -400 V). The SiC detector is divided in 4 pixels, each of them with a surface of $\approx$ 5$\times$5 mm$^{2}$, with a common connection to the ground.
 A Mesytec preamplifier (MPR-16) and a CAEN (DT5742) digitizer, set with a frequency of 1 GHz, 14-bit, have been used for such measurements \cite{ref:mesytech, ref:digitizer_caen}.\\
A characterization of the SiC detector was performed using a mixed nuclide $\alpha$-source (Am, Pu, Cm), with the aim of investigating the SiC performances in terms of energy and timing resolutions, maximizing the signal-to-noise ratio. Figure \ref{Picture_SiC_detector} (left) shows a picture of the 2$\times$2 pixels SiC detector, while Figure \ref{Picture_SiC_detector} (right) evidences the $\alpha$ mixed source placed in front of the detector.
Using the CAEN digitizer, the signal waveforms have been acquired and digitized signals have been filtered to reconstruct the energy, the detection time, the rise and fall times and other relevant parameters, as discussed in detail in ref. \cite{ref:CARDELLA2024169961}. As an example of the acquired data, Figure \ref{figure_waveforms} --red line-- shows the waveform of a signal produced by an $\alpha$ particle, obtained from a SiC pixel, while the black line represents the signal after the applied trapezoidal filter, which allows for extracting the maximum and provides information about the energy. Furthermore, a triangular filter, able to perform a smoothing of the signal waveform is used to extract information about the rise-time and the starting time of the preamplifier output signal \cite{ref:JORDANOV1994337}. In detail, the rise-time is evaluated as the time difference between the 30\% and the 80\% level of the signal with respect to the maximum. The start or crossing time is evaluated as the time in which the signal overcomes a threshold, evaluated as a defined percentage of the maximum, as in the case of direct constant fraction discrimination discussed in ref. \cite{ref:Amorini_4484233}. In both cases the maximum is the one calculated with the trapezoidal filter.\\ 
Figure \ref{energy-risetime-spectra} (a) shows the calibrated energy vs rise-time plot, obtained for a single SiC pixel. As it can be noticed, there are some events with an average rise-time $\gtrsim$ 7 ns and lower energy. These events are mainly ($\gtrsim$ 90\%) associated with multiplicity, i.e. number of hit pixels in the same event, M$=$1, while a lower percentage is due to events with M$=$2. The events with rise-time $\gtrsim$ 7 ns and lower energy, indicated with a black ellipse in Figure \ref{energy-risetime-spectra} are due to an edge effect. As usual in many detectors, the charge collection is worse around the side region than the center of the detector, as it will be discussed in detail in the next section. Figure \ref{energy-risetime-spectra} (b) shows the projected calibrated energy spectrum, obtained for the single SiC pixel. This spectrum has been obtained with a cut in M$=$1 and in rise-time $\lesssim$ 7 ns. Performing gaussian fits, an energy resolution of $\approx$ 50 keV (FWHM) has been obtained. 
A further test to study the energy resolution has been performed using an ORTEC 542 amplifier and a Multi Channel Analyzer (MCA), as shown in Figure \ref{MCA-spectrum}. Using the ORTEC amplifier and optimizing the shaping time, a resolution (FWHM) of $\approx$ 30 keV (0.5\%) has been obtained. A similar resolution has been obtained for a Si detector with a surface of 1.7 cm$^{2}$ and a thickness of $\approx$ 100 $\mu$m.
Energy resolution is improved when employing the amplifier compared to tests conducted alone with a digitizer; this is most likely because the amplifier shaping increases noise rejection. Furthermore, a study on the digital filter is ongoing to improve the energy resolution. 
\subsection{Investigation of the edge effect}
\begin{figure}
\centering
\includegraphics[scale=0.50]{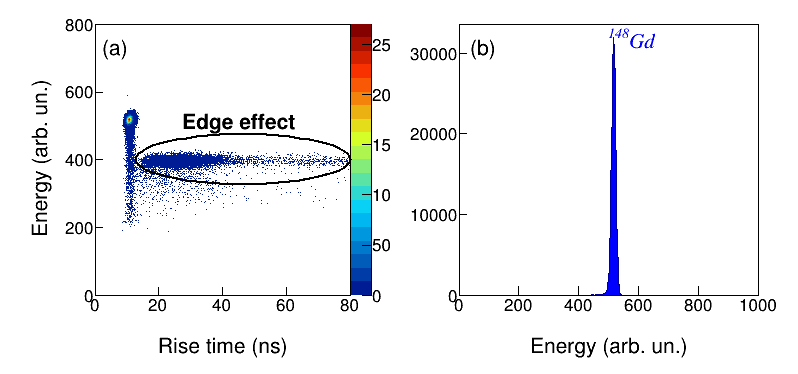}
\caption{(a) Energy vs rise-time plot obtained using a $^{148}$Gd $\alpha$-source for a SiC pixel of 10 $\mu$m. (b) Energy spectrum obtained as a projection in the x axis of figure shown in the (a) panel with a cut in M$=$1 and in rise-time$\lesssim$14 ns. \label{energy-risetime-spectra-Gado}}
\end{figure}

\begin{figure}
\centering
\includegraphics[scale=0.45]{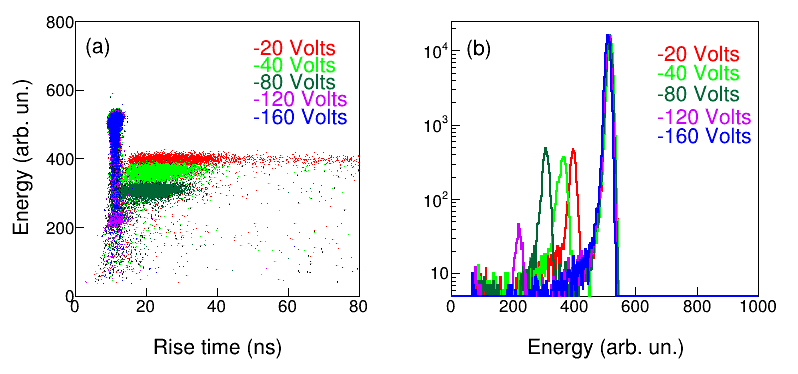}
\caption{(a) Energy vs rise-time plot obtained for several voltages applied to the 10 $\mu$m detector, as reported in the legend. (b) Energy spectrum obtained for several voltages, as reported in the legend. \label{spectra-several-voltage}}
\end{figure}

A test to better understand the presence of events with higher rise-time and lower energy has been conducted using a 2$\times$2 pixels 10 $\mu$m thick-1 cm$^{2}$ detector and a $^{148}$Gd $\alpha$-source. This radioactive source was used because the $\alpha$ particles emitted are stopped in the detector. 
In fact, with a 10 $\mu$m detector, it is possible to study the edge effects operating in safe conditions at much higher voltages than the nominal depletion one (-20 $V$). Which is not the case for the 100 $\mu$m detector, indeed reaching very high depletion voltage ($>$ -500 $V$) could damage the used electronics.
Figure \ref{energy-risetime-spectra-Gado} (a) shows the uncalibrated energy vs rise-time plot, obtained for a single SiC pixel. As in the case of the 100 $\mu$m SiC, some events with a higher rise-time ($\gtrsim$ 14 ns) and a lower energy are present. These events are mainly ($\gtrsim$ 90\%) associated with M$=$1, while a lower percentage is due to events with M$=$2. The increase of the rise-time compared to the 100 $\mu$m is due to the capacity effect, that in this case is higher than the 100 $\mu$m SiC. Figure \ref{energy-risetime-spectra-Gado} (b) shows the energy spectrum, obtained for the same SiC pixel, with a cut in M$=$1 and in rise-time $\lesssim$ 14 ns. Performing a gaussian fit, an energy resolution of $\approx$ 3$\%$ FWHM has been obtained.  
Figures \ref{spectra-several-voltage} (a),(b) show the results obtained changing the voltage from -20 $V$ up to -180 $V$.  As it can be observed for a SiC pixel, at the increasing of the voltage the presence of the region with higher rise-time decreases, confirming that this effect is due to a worse charge collection in the edge of the detector. This observation has been further verified by a measurement using a collimator positioned in front of the central region of the SiC pixel, which has shown that the presence of events with higher rise time is almost completely removed.\\
It is important to note that the guard ring of the detector is floating. Consequently, increasing the depletion voltage of the detector also alters the effective potential of the guard ring. This variation likely enhances the charge collection efficiency of the guard ring. Such behavior may account for the observed suppression of the edge effect with increasing voltage.\\ 
In conclusion of this investigation, it is remarkable that the detector withstood an 800\% overvoltage without exhibiting any signs of electrical breakdown, thereby demonstrating the robustness of the device. By contrast, a silicon-based detector subjected to a comparable overvoltage would likely have suffered irreversible damage. Ongoing simulations are being conducted to gain a comprehensive understanding of the mechanisms underlying the edge effect \cite{ref:Dagata_IWM2024}.
\section{Timing investigation: algorithms and sharing method data analysis}
\label{Timing_investigation}
To obtain the timing information, it is necessary to collect a precise timing evaluation of one point of the acquired waveform for each signal. To this purpose, we used a Threshold Crossing-Time (TCT) algorithm, able to determine the time at which the waveform crosses a chosen threshold level. The algorithm is based on a polynomial fitting, as described in detail in ref. \cite{ref:Amorini_4484233}. In the first step, the algorithm scans the waveform to estimate the crossing-sample. In order to increase the precision, in the second step the crossing-sample is used as the centroid of an 8-sample window used to apply a least-square cubic fitting. Finally, a bisection method performs more precise threshold crossing-time estimation. In this way, as described in detail in ref. \cite{ref:Amorini_4484233}, a better time-resolution  than the one obtained with the sampling time can be reached.\\ 
To obtain the information on the timing performance of a single SiC pixel, we carried out data analysis looking at the sharing of the signals; namely signals in events in which two adjacent pixels are in coincidence.  This usually occurs when a particle impinges on the interpixel region. Some very preliminary results, obtained in air, are reported in ref. \cite{ref:Martorana_IWM2024}. In the following, results obtained using this method with a mixed $\alpha$ source, a $^{148}$Gd $\alpha$ source and low-energy protons and $\alpha$ beams are discussed.\\ 
\subsection{Timing investigation using radioactive $\alpha$ sources}
\begin{figure}
\centering
\includegraphics[scale=0.40]{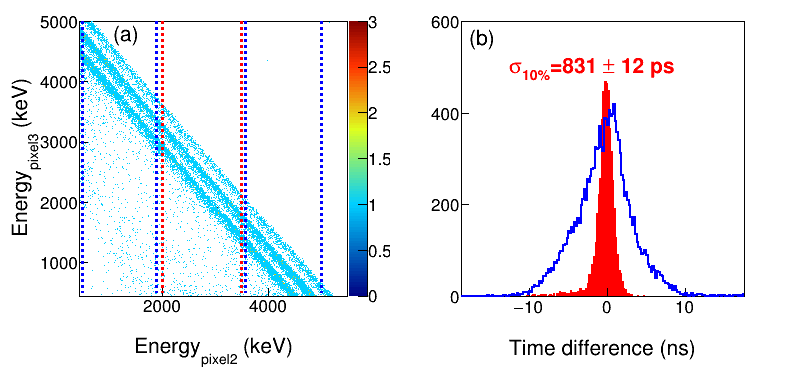} 
\caption{(a) Energy of a pixel versus the energy of an adjacent pixel in coincidence events. (b) Timing difference between the crossing signals in coincidences events. The red distribution is obtained with a cut in energy (2000 keV$<$ $E_{pixel_{2}}$ $<$ 3500 keV), the blue distribution is obtained with cuts in energy of 500 keV$<$ $E_{pixel_{2}}$ $<$ 2000 keV and of 3500 keV$<$ $E_{pixel_{2}}$ $<$ 5000 keV. The $\sigma$ value has been obtained performing a gaussian fit.}\label{coincidence_events}
\end{figure}

\begin{figure}
\centering
\includegraphics[scale=0.25]{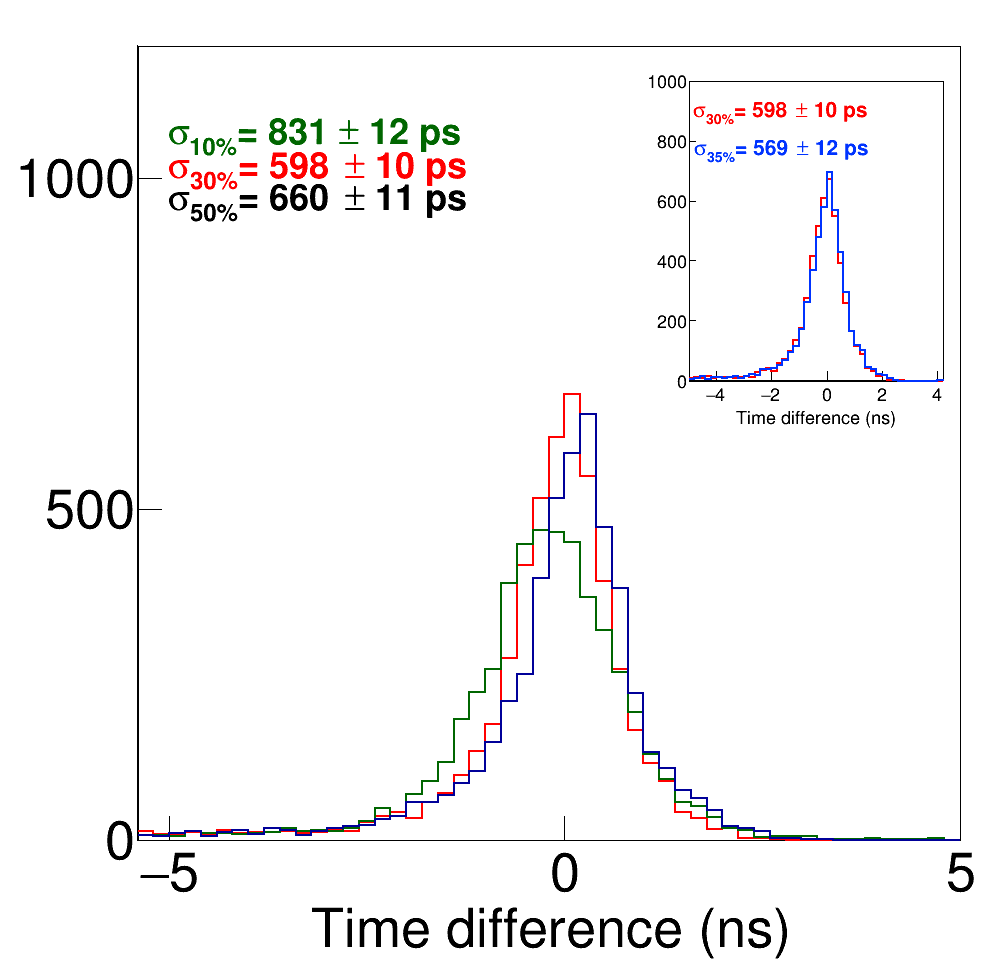}
\caption{Timing differences between the start signals, obtained with the cuts shown in Figure \ref{coincidence_events} (a) as red dotted lines. Different colours are referred to different crossing times, as shown in the legend. The $\sigma$ values have been obtained within gaussian fits. The upper panel shows the timing distributions for the crossing time of 30\% and 35\%. \label{Figure_different_timing}}
\end{figure}

\begin{figure}
\centering
\includegraphics[scale=0.45]{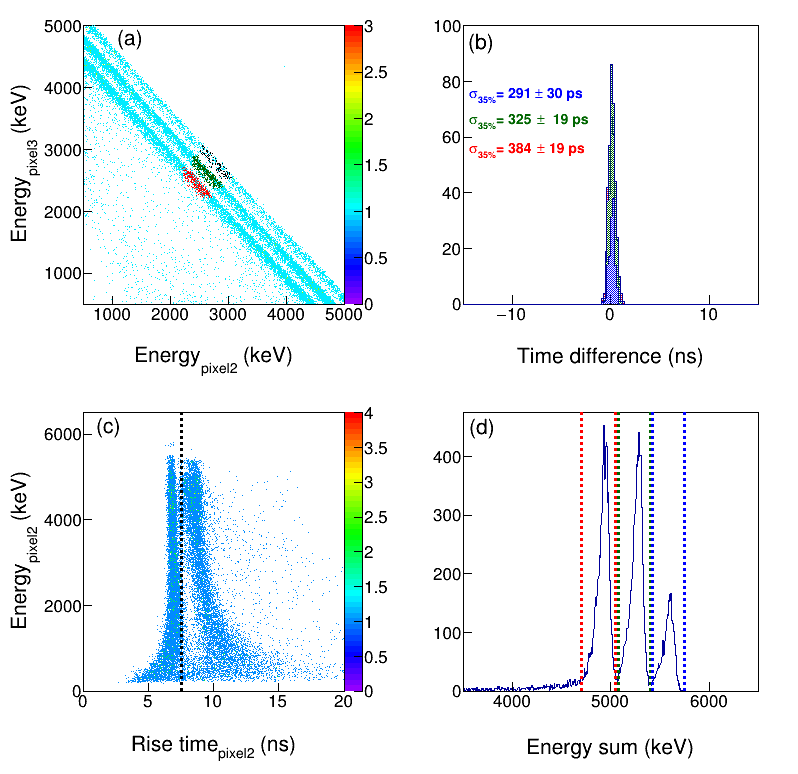} 
\caption{(a) Energy of a pixel versus the energy of an adjacent pixel in coincidence events. 
(b) Timing difference between the crossing time signals obtained with cuts shown in Figures (a,c,d).
(c) Energy vs rise-time plot of a pixel in coincidence events with the adjacent pixel. (d) Energies sum of two pixels.  $\sigma$ have been obtained within gaussian fits.\label{results_interpad_event_selection_mixed_alpha}}
\end{figure}

The first measurement has been conducted in vacuum, placing the $\alpha$ mixed source in front of the SiC 100 $\mu$m 2$\times$2 pixels detector. Figure \ref{coincidence_events} (a) shows the energy of a pixel (number 2) versus the energy of an adjacent pixel (number 3) in coincidence events. Three lines are visible, roughly corresponding to the three energies of the $\alpha$ mixed source. Figure \ref{coincidence_events} (b) shows the time difference between the start/crossing time signals of the two pixels (number $2$ and $3$) in coincidence. As first step, the start/crossing time has been evaluated at the 10\% of the extracted maximum, following the aforementioned TCT method.\\ In detail, Figure \ref{coincidence_events} (b) displays two timing distributions: the red one is obtained including a cut in the central energy region of pixel$2$ (2000 keV $<$ $E_{pixel2}$ $<$ 3500 keV)-- shown as red dotted lines in Figure \ref{coincidence_events} (a); the blue curve is obtained with cuts, shown as blue dotted lines in Figure \ref{coincidence_events} (a) (500 keV$<$ $E_{pixel2}$ $<$ 2000 keV and 3500 keV$<$ $E_{pixel2}$ $<$ 5500 keV), where the energy released in a pixel is lower than the released energy of the adjacent pixel. The observed trend is in agreement with what is expected. Indeed, the timing distribution worsens in the case in which the energy of a pixel is lower than the energy of adjacent pixels. This is because the timing resolution is more affected by the pixel where lower energy is released. For the red timing distribution, a $\sigma$=831 $\pm$ 12 ps has been extracted performing a gaussian fit.\\
By following this procedure, a data analysis was conducted with the crossing time varied to determine the value that permits the lower $\sigma$. Figure \ref{Figure_different_timing} shows the timing distributions obtained using different crossing time, i.e. 10$\%$, 30$\%$, 50$\%$; such timing distributions are obtained with the energy cuts shown as red dotted lines in Figure \ref{coincidence_events} (a). The upper inset of Figure \ref{Figure_different_timing} shows the timing distributions obtained using the crossing time of 30$\%$ and 35$\%$, respectively. 
As it can be observed, the lower $\sigma$ is obtained within a crossing time of 35$\%$. This result is similar to the one obtained in ref. \cite{ref:WANG2022166050}, with the equivalent direct constant fraction discrimination method.\\ In the following, due to previous results, the value of 35\% crossing time will be used to obtain the timing information. Using this crossing time, to extract more information about the timing performance of a single pixel, we focused the data analysis on signals with approximately the same energy in both adjacent pixels (2000 keV$<$ $E_{pixel_{2}}$ $<$ 3500 keV). 
In order to better characterize the energy dependence of the time resolution, more strict conditions were included in the analysis: i) a rise-time$<$7.5 ns in order to remove edge-effects due to the guard ring events; ii) a costraint upon the sum of the two pixels energies; iii) a cut on the energies of two adjacent pixels (k,j) defined by the following expression:
\begin{equation}
\frac{\vert(E_{pixelk}-E_{pixelj})\vert}{\frac{(E_{pixelk}+E_{pixelj})}{2}}<0.2 
\end{equation} 
where the $E_{pixelk}+E_{pixelj}$ has been evaluated by performing gaussian fits on the peaks shown in Figure \ref{results_interpad_event_selection_mixed_alpha} (d).
Figure \ref{results_interpad_event_selection_mixed_alpha} (b) shows the timing distributions obtained including such cuts, shown as dotted lines in Figures \ref{results_interpad_event_selection_mixed_alpha} (c) and (d), and as red, blue and green marker color in Figures \ref{results_interpad_event_selection_mixed_alpha} (a).
The cuts in energy allow to select particles loosing almost equal energies in the two pixels, with a 10\%  discrepancy in energy. Within this data analysis an average $\sigma$ of $\approx$ 340 ps has been obtained, with a slight dependence on the sharing energy.\\ Supposing that the time response of the two adjacent pixels is the same for a similar energy range, the total $\sigma_{tot}$, i.e. the value extracted performing gaussian fits, is given by: 
\begin{equation}
\sigma_{tot}=\sqrt{(\sigma_{pixel_{k}})^2+(\sigma_{pixel_{j}})^2}
\end{equation}
\begin{equation}
\sigma_{tot}=\sqrt2\sigma_{pixel_{k;j}}
\end{equation}
\begin{equation}
\sigma_{pixel_{k;j}}=\frac{\sigma_{tot}}{\sqrt2}
\end{equation}
Following this assumption, an average $\sigma_{pixel}$ $\approx$ 240 ps has been obtained. 
The same data analysis has been carried-out using a $^{148}$Gd $\alpha$ source. In this case, a $\sigma_{tot}$= 679 $\pm$ 107 ps ($\sigma_{pixel}$ $\approx$ 490 ps) has been obtained in the energy region of $\approx$ 1400 keV.\\
It should be noted that the sum of the energies of two adjacent pixels (Figure \ref{results_interpad_event_selection_mixed_alpha} (d)) is similar to the peak energy value (Figure \ref{energy-risetime-spectra} (b)), with a loss in energy of the $\approx$ 5\%, due to the fact that some losses in charge can happen. This is particularly important for the development of a detector array, as event losses due to interpixel effects can impact detection efficiency, as discussed in more detail in ref. \cite{ref:CARDELLA2024169961}.
\subsection{Timing investigation with $\alpha$ and proton beams}
\begin{figure}
\centering
\includegraphics[scale=0.40]{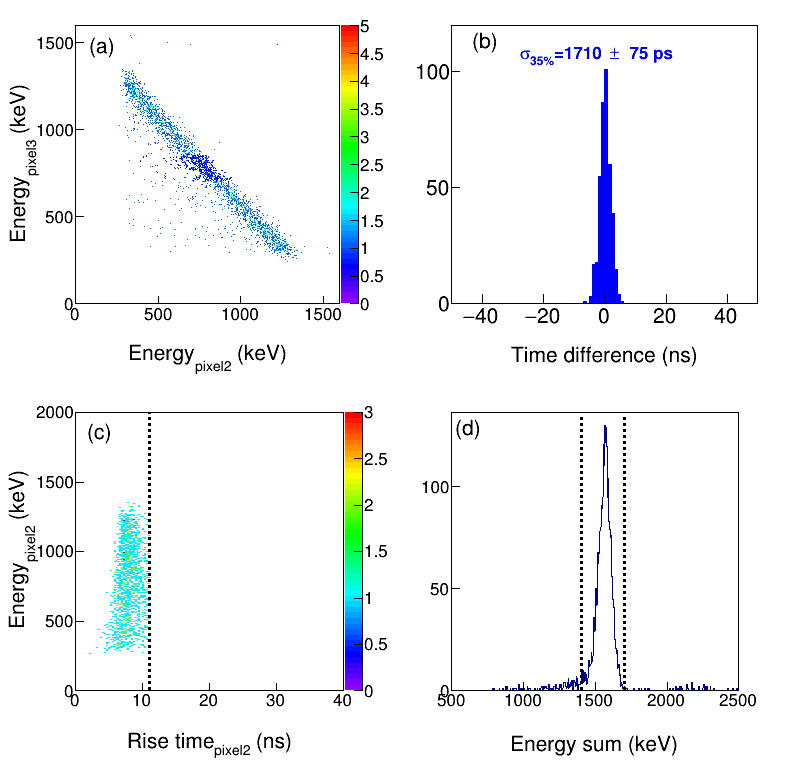}
\caption{(a) Energy of a pixel versus the energy of an adjacent pixel in coincidence events, with the 2 MeV $\alpha$ beam. (b) Timing difference between the start signals, obtained with cuts shown in Figures (a,c,d), $\sigma$ has been obtained within a gaussian fit. (c) Energy vs rise time plot of a pixel in coincidence events with the adjacent pixel. (d) Energies sum of two pixels.  
\label{coincidence_events_alpha}}
\end{figure}

\begin{figure}
\centering
\includegraphics[scale=0.20]{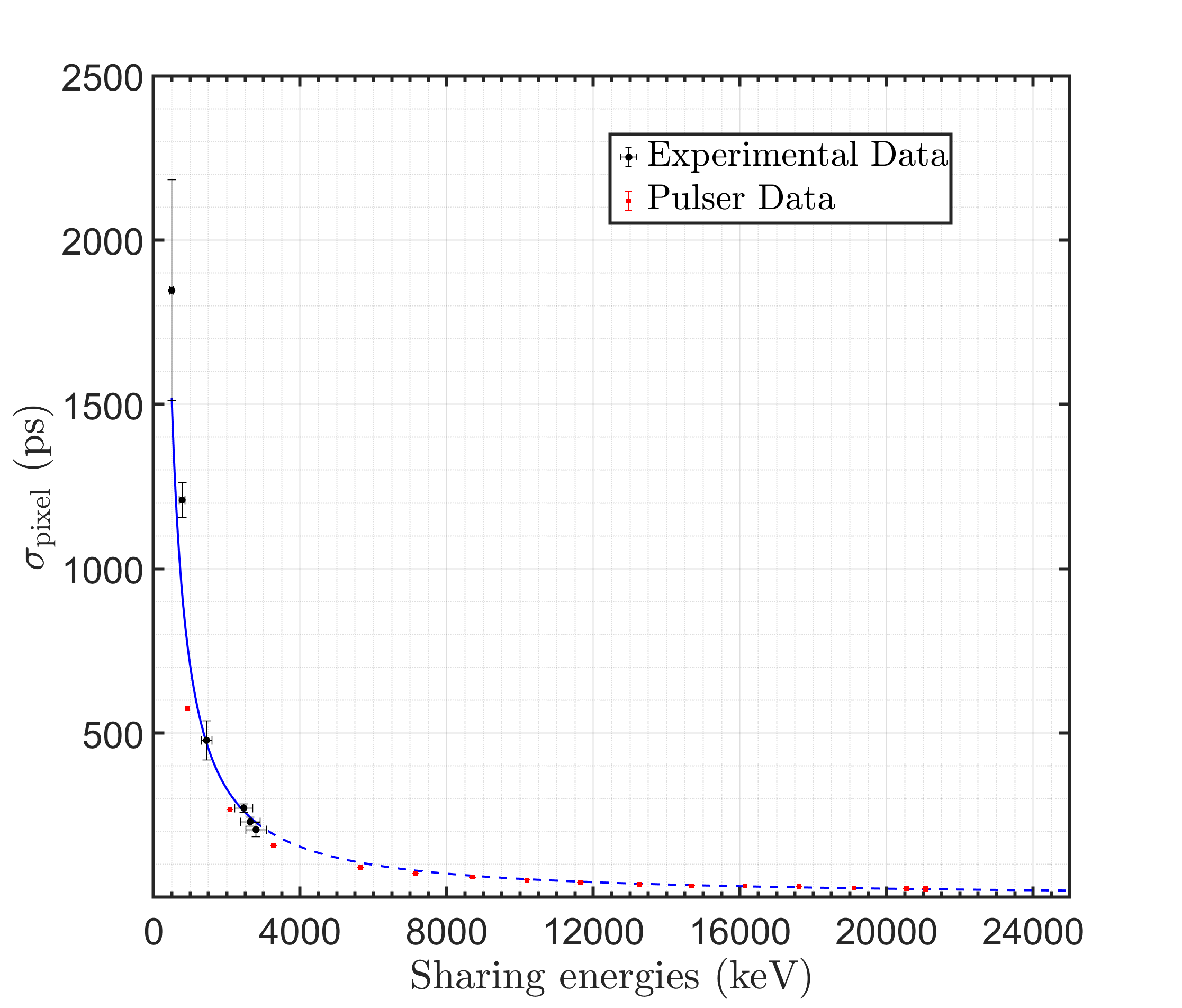}
\caption{$\sigma_{pixel}$ values as a function of the sharing energies. \label{trend_sigma} Errors bars are statistical on y axis and due to the selected energy region (10\%) on the x one. Black points: experimental data. Blue curve: fit and extrapolation at higher energies. Red points: data obtained with a CAEN emulator.}
\end{figure}

Finally, we performed a test using low-energy proton and $\alpha$ beams provided by the singletron accelerator of the Dipartimento di Fisica and Astronomia (DFA) \textit{Ettore Majorana}, of the University of Catania (Italy). The singletron accelerator provided protons with energy of 1 MeV and $\alpha$ with energies of 1 and 2 MeV.\\ The measurement has been performed in a backscattering mode, placing the 100 $\mu$m SiC detector at $\theta$ $\approx$ 150$^\circ$ in the laboratory frame, inside a vacuum chamber. A thick carbon target with an evaporated Au layer $\approx$ 20 $\mu$g$/$cm$^2$ was used. The thick carbon substrate was used to reduce the low energy backscattering due to the steel target holder. 
Figure \ref{coincidence_events_alpha} displays the results obtained using the sharing method and the procedure discussed before, for $\alpha$ beam at 2 MeV. For proton beam at 1 MeV a $\sigma$=2709 $\pm$ 450 ps is obtained. While, in the case of $\alpha$ beam at 1 MeV the shared signals are too low in energy to be detected.\\ 
To conclude this timing investigation, Figure \ref{trend_sigma} shows the trend of the extracted $\sigma_{pixel}$ (ps) values as a function of the sharing energies. This trend roughly goes as $\approx$ 1/E, in agreement with what expected.
A fit has also been performed and the extrapolation at a value of 10 MeV shows a $\sigma_{pixel}$ of $\approx$ 50 ps. In addition, using a CAEN signal emulator, signals with a rise-time similar to the one observed with $\alpha$ sources  and several amplitudes, from 30 mV up to 800 mV, have been sent to the test input of the Mesytec preamplifier, in order to test the intrinsic timing resolution of the preamlifier coupled with the CAEN digitizer. The values, showed as red points in Figure \ref{trend_sigma},  show that the used electronics is the most important limitation to the extracted time resolution at high energy.\\ We underline that the figures and procedure have been discussed for two adjacent pixels, but this trend is also confirmed for the remaining adjacent pixels.
In conclusion, we were able to extract several information on the timing resolution as a function of the sharing energy. These information were extracted using $\alpha$ radioactive sources and low-energy beams. Furthermore, an analysis on the crossing time allowed to find the value to obtain the optimal timing resolution. Future research and development of TCT algorithms may provide an improvement in the measured time resolution \cite{ref:Amorini_4484233, ref:WANG2022166050}.
\section{Timing performance using a Micro-channel Plate (MCP) detector}
\begin{figure}
\centering
\includegraphics[scale=0.20]{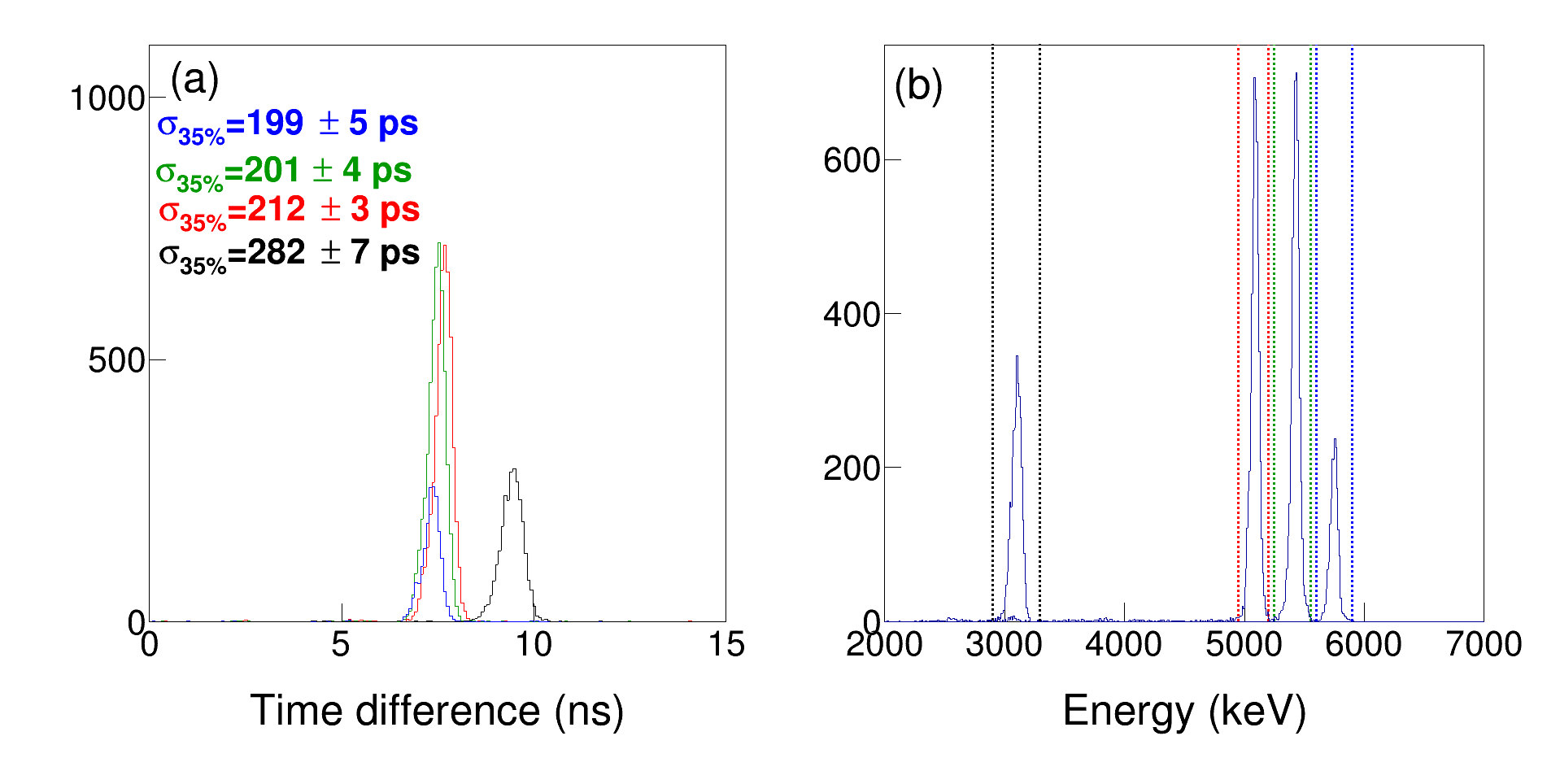}
\caption{(a) Time differences between the start signals of the MCP and a pixel of the 2$\times$2 SiC array, using a frequency of 1 GHz, and two $\alpha$ sources (mixed and $^{148}$Gd one). Different timing distributions are in coincidence with the energy spectra, shown in panel (b). (b) Energy spectra measured with a pixel of SiC.}\label{Figure_Results_MCP+SiC_mixedGd}
\end{figure}

In order to confirm the obtained results and extract the timing information, a test using the micro-channel plate (MCP) detector, described in ref. \cite{ref:MUSUMARRA2010399}, as start signal and the 100 $\mu$m 2$\times$2 pixels SiC detector as stop has been performed, over a base of flight of $\approx$ 11 cm. The test was carried out at INFN-LNS, in a vacuum chamber using a mixed and a $^{148}$Gd $\alpha$ sources. For the SiC detector the standard Mesytec preamplifier was used, while for the MCP a fast preamplifer was used.
The data acquisition was triggered by the OR of the four pixels of SiC detector. In this configuration, the crossing time has been determinated as the 35$\%$ of the maximum of the waveform. Figure \ref{Figure_Results_MCP+SiC_mixedGd} shows the results obtained using the MCP+SiC detectors and the two radioactive sources. In Figure \ref{Figure_Results_MCP+SiC_mixedGd} also the obtained $\sigma$ values are reported. 
The MCP + SiC measurement allowed to test the timing performance, confirming that a $\sigma$ of $\approx$ 300 ps is achieved in the 3000 keV energy region. This result corroborates the findings obtained using the sharing method. Taking into account the contribution from the MCP resolution, not measured, and the time spread of the particle trajectories due to the MCP foil tilted at 45$^\circ$, that can be evaluated to be up to $\approx$ 100 ps, this value is quite similar to the resolution measured at $\approx$ 2500 keV with the method of charge sharing between the two pixels. Additionally, it provides the opportunity to evaluate the timing performance at slightly higher energies, such as around 5000 keV, showing a value of $\approx$ 200 ps. This value is worse than the value expected within the extrapolation reported in Figure \ref{trend_sigma} (100 ps at 5000 keV). This difference can arise from the MCP contribution and from geometrical effects. 
\section{Conclusions}
The characterization of 2$\times$2 pixels-- 1 cm$^{2}$, 100 $\mu$m-- SiC detector has been presented in this paper, with a focus on its energy and timing resolutions for potential applications in medical and nuclear fields, specifically using radioactive ion beams. The study, conducted using commercial electronics, radioactive sources, and low-energy proton and $\alpha$ beams, shows an energy resolution $\approx$ 1\% in the 5000 keV energy range and a timing resolution of approximately 200 ps ($\sigma_{pixel}$), within the 2000–3000 keV energy range. This timing resolution was determinated by using the sharing method, based on the crossing time difference of coincidence signals, allowing for precise timing performance assessment of segmented detectors. This method enables the extraction of individual pixel or strip timing resolution and it has been validated using a micro-channel plate detector as time reference. Furthermore, tests on the Time Crossing Threshold algorithm, varying the crossing time values have permitted to identify the optimal crossing time, minimizing the $\sigma$ values, i.e. obtaining the optimal time resolution.  
The promising results highlight the potential of these detectors for high-performance energy and time measurements in various applications. Future work will concern the use of ion beams, as carbon, oxygens beams, of higher energy, allowing also to explore the timing as a function of the fragments energy, not only to extend the investigation at higher energy but also to verify eventual dependence from the charge of the ion in order to check pulse shape effects.
\section{Acknowledgment}
This work has been partially funded by European Union (NextGeneration EU),
through MUR-PNRR project SAMOTHRACE (ECS00000022), PIF2024-PID2023-147569NB-C21 and
MRR-ISRS-SPAIN CIE2301005. Authors are grateful to the Servizio tecnologie avanzate of INFN-Sezione di Catania, and in particular to N. Giudice, for the contribution on the bonding of detectors. 



\bibliographystyle{elsarticle-num-names} 
\bibliography{Biblio-bib.bib}


%
%
%
%
\end{document}